\def\year{2022}\relax
\documentclass[letterpaper]{article} 
\usepackage{aaai21}  
\usepackage{times}  
\usepackage{helvet} 
\usepackage{courier}  
\usepackage[hyphens]{url}  
\usepackage{graphicx} 
\usepackage{natbib}  
\usepackage{caption} 
\usepackage{enumitem}
\usepackage{tabularx}
\usepackage{enumitem}
\usepackage{amssymb} 
\usepackage{amsmath} 
\usepackage{amssymb} 
\usepackage{txfonts} 
\usepackage{subfig}
\usepackage[table]{xcolor} 
\usepackage{siunitx} 
															  
\usepackage{scrextend}
\deffootnote[2em]{2em}{1em}{\textsuperscript{\thefootnotemark}\,}

\usepackage{pgfplots}
\usepackage{pgfplotstable}
\usepgfplotslibrary[groupplots]
\usetikzlibrary{plotmarks}

\usepackage{xcolor}
\definecolor{mypurple}{rgb}{0.8, 0.502, 1}
\definecolor{myblue}{rgb}{0.302, 0.4745, 1}
\definecolor{myred}{rgb}{1,0,0}

\usepackage{arydshln}
\usepackage{multirow}   
\usepackage{footnote}   
\usepackage[multiple]{footmisc} 
\usetikzlibrary{positioning,shapes,arrows,shadows,patterns} 

\newcommand{\ignore}[1]{}
\usepackage[show]{notes}
\newcommand{\com}[1]{}

\usepackage[normalem]{ulem}

\pgfplotsset{compat=1.11,
        /pgfplots/ybar legend/.style={
        /pgfplots/legend image code/.code={%
        \draw[##1,/tikz/.cd,bar width=3pt,yshift=-0.2em]
                plot coordinates {(0cm,0.8em)};
        },
    },
}

\title{Going Extreme: Comparative Analysis of Hate Speech in Parler and Gab}

        \author {
            Abraham Israeli~~~~~~~~~~~~
            Oren Tsur \\ 
        }
        \affiliations {
            Department of Software and Information System Engineering \\ Ben-Gurion University of the Negev \\ Beer-Sheva, Israel\\
            isabrah@post.bgu.ac.il~~~~~~~~~~~~orentsur@bgu.ac.il
        }

\begin{document}

\maketitle    

\begin{abstract}
Social platforms such as Gab and Parler, branded as `free-speech' networks, have seen a significant growth of their user base in recent years. This popularity is mainly attributed to the stricter moderation enforced by mainstream platforms such as Twitter, Facebook, and Reddit.
In this work we provide the first large scale analysis of hate-speech on Parler.

We experiment with an array of algorithms for hate-speech detection, demonstrating limitations of transfer learning in that domain, given the illusive and ever changing nature of the ways hate-speech is delivered. In order to improve classification accuracy we annotated 10K Parler posts, which we use to fine-tune a BERT classifier. Classification of individual posts is then leveraged for the classification of millions of users via label propagation over the social network. Classifying users by their propensity to disseminate hate, we find that hate mongers make 16.1\% of Parler active users, and that they have distinct characteristics comparing to other user groups. We find that hate mongers are more active, more central and express distinct levels of sentiment and convey a distinct array of emotions like anger and sadness.
We further complement our analysis by comparing the trends discovered in Parler and those found in Gab. 

To the best of our knowledge, this is among the first works to analyze hate speech in Parler in a quantitative manner and on the user level, and the first annotated dataset to be made available to the community. 

\end{abstract}

\section{Introduction}
\label{sec:intro}

[{\bf Warning}: Some of the readers may find the language in the examples provided in this manuscript offensive.]\\

Social platforms like Twitter, Facebook, and Reddit have become a central communication channel for billions of users\footnote{Facebook reported on 2.9 Billion monthly active users (retrieved 07/28/2021), see: https://tinyurl.com/2p8r4wd6}. However, the immense popularity of social platforms resulted in a significant rise in the  toxicity of the discourse, ranging from cyber-bullying to explicit hate speech and calls for violence against individuals and groups \cite{waseem2016hateful,mondal2017measurement,laub2019hateGlobalComparison,ziems2020racism}. Women, people of color, the LGBT community, Muslims, immigrants, and Jews are among the most targeted groups. Recent studies report on a surge in Islamophobia \cite{akbarzadeh2016muslim,sunar2017long,osman2017retraction,chandra2021virus}, antisemitism \cite{adl2020antisemitism,zannettou2020quantitative}, xenophobia \cite{iwama2018understanding,entorf2019refugees}, hate of Asians \cite{an2021predicting,vidgen2020detecting} and hate crimes \cite{dodd2017anti,levin2018report,edwards2018effect,perry2020planting}.

\begin{table*}[ht!]
\footnotesize
\centering
\begin{tabular}{c|m{13.5cm}|c}
\hline
\rowcolor{gray!40} 
& \multicolumn{1}{c}{Post Content} & Target\\ \hline\hline
1 & \texttt{Raimondo can stick her mask mandate in her cunt .come tell me to my face it will be the last time we here from this bitch}  &  Women \\ \hline
2 & \texttt{WDammit I hate kikes so much. Filthy, hooked nosed, baby penis sucking parasites} & Jewish people \\ \hline
3 & \texttt{Muslim terrorists did this and obama brought them here to run our government. Gotta get rid of Obammy, Omar and all these other fucking Muslims}   & Muslims \\ \hline
4 & \texttt{Remember what (((they))) want to take from you} & Jewish People \\ \hline


5 & \texttt{SMASH these radical freaks all over AMERICA !! Clean this mess up } & LGBTQ \\ \hline

6 & \texttt{The CHINA plan is for biden to bring millions and millions of CHINA virus infected illegal savages into america.} & Asians, Immigrants \\ \hline

7 & \texttt{Now the blacks not only want welfare, free college, free healthcare, free housing, free food, free clothes, free transportation, free tampons, but they want the white man money. Wake up and smell the dog shit.} & The Black Community \\ \hline

8 & \texttt{DemonKKKrats love rape and murder. Praying.} & Democrats \\ \hline
\end{tabular}
\caption{A sample of posts from Parler social platform.}
\label{table:posts_sample}
\end{table*}

Facing an increased public and legislature scrutiny, mainstream social platforms (e.g., Facebook, Twitter, Reddit) committed to a stricter enforcement of community standards, curbing levels of hate on the platform\footnote{https://tinyurl.com/muvn4hma}\footnote{https://tinyurl.com/yc3kx6wp}.

The stricter moderation of content drove many users into joining alternative social platforms such as Parler
and Gab. Touting their commitment to `free speech' and `no moderation' policy, these platforms attract users suspended from mainstream platforms, conspiracy theorists, extremists and other unhinged users, as well as `free-speech' advocates. 

User migration to Parler and Gab was not only grass-root. The platforms were promoted by prominent news anchors and political figures. For example, U.S. Senator Ted Cruz (R-TX) tweeted \emph{\small ``I’m proud to join @parler\_app -- a platform gets what free speech is all about -- and I’m excited to be a part of it. Let’s speak. Let’s speak freely. And let’s end the Silicon Valley censorship''} (6/25/2020), and Sean Hannity, a popular host and commentator on Fox news, informed the viewers of his daily show that \emph{\small ``I saw that the president had joined it. At least there is a place, it's like Twitter, it's called Parler, I have an account there... good for you because the president joined, because they are censoring him and Dan Scavino and everybody else''} (1/8/2021).

Hate, brewing online, often spills to the streets \cite{splc2018altright,altrightpipeline,violence2019,8chan}. Thus, defending `hate speech' under the right for `free speech' may manifest itself through very concrete actions in ``real life''.  The perpetrator of the Pittsburgh synagogue shooting\footnote{\url{https://tinyurl.com/6rpn5j67}} was active on Gab, referring to ``the infestations of jews''. His final post, minutes before opening fire in the synagogue, was \emph{\small ``I can't sit by and watch my people get slaughtered. Screw your optics, I'm going in.''} Similarly, the storming of the U.S. Capitol on January 6, 2021 was found by the U.S. Senate investigation committee to be encouraged and coordinated on Parler \cite{capitol2021}. 

Indeed, hate speech does plague Parler -- a number of examples is presented in Table \ref{table:posts_sample}. Notably, some posts are more explicit than others -- using vulgar language (e.g., posts \#1--\#3), explicitly mentioning the targeted individual/group (e.g., \#1,\#3,\#6,\#7), while other posts are using nick-names, codes and implicit references (e.g., \#2,\#4,\#5,\#8).

Striking the right balance between contradicting values (e.g., the freedom of speech vs. public safety of members of protected groups) is a walk on a tightrope. We believe, however, that a data-oriented analysis may help individuals and policy maker alike at reaching an informed balance.

In this work we focus on Parler social platform, investigating the proliferation of hate speech on the platform, both on the post level and on the user level. We identify three distinct groups of users (hate mongers, regular users and hate flirts) and show significant differences between them in terms of language, emotion, activity level and role in the network. We further compare our result to the hateful dynamics observed in the Gab platform.  
\paragraph{Contribution} Our contribution in this paper is fourfold: {\bf (i)} We compare an array of state-of-the-art algorithms for hate detection, showing they all fail to accurately identify nuanced and novel manifestations of hate speech found on Parler, {\bf (ii)} We share the first annotated Parler dataset, containing 10K Parler posts, each post labeled by the level of hate it conveys, {\bf (iii)} We fine-tune a BERT-based classifier to achieve accurate classification, and modify DeGroot’s  diffusion model \cite{golub2010naive} in order to allow analysis on the platform level, and finally {\bf (iv)} We provide the first large scale analysis of the proliferation of hate in Parler and compare it to the user dynamics in Gab.  
\\\\
The remainder of the paper is organized as follows: Section \ref{sec:related_work} provides a brief review of the relevant literature. A detailed description of the datasets and the annotation procedure are given in Section \ref{sec:data}. In Section \ref{sec:methods} we present the computational methods we use for the post and user level classification, and results follow in Section \ref{sec:results}. 
A detailed analysis of hate levels and user propensity for hate speech in Parler and Gab is provided in Section \ref{sec:analysis}. Finally, Section \ref{sec:discussion} offers some discussion regarding some of the observations, including ethical considerations.


\section{Related Work}
\label{sec:related_work}
A growing body of research studies the magnitude and the different manifestations of hate speech in social media \cite{knuttila2011user, chandrasekharan2017you, zannettou2018gab, zampieri2020semeval, ranasinghe2020multilingual}, among others. Here, we present an overview of the current literature through three different perspectives: (i) The detection of hate speech on the \emph{post} level, (ii) The detection of hate-promoting \emph{users}, and (iii) The characterization of hate speech on the \emph{platform} level.

\paragraph{Post-level classification} Most previous works address the detection of hate in textual form. Keywords and sentence structure in Twitter and Whisper were used in \cite{mondal2017measurement,saleem2017web}, demonstrating the limitations of a lexical approach. 

The use of code words, ambiguity and dog-whistling, and the challenges they introduce to text-based models  were studied by \cite{davidson2017automated,ribeiro2017like,arviv2021sa}.
The detection of implicit forms of hate speech is addressed by \citet{magu2017detecting} which detects the use of hate code words (e.g., google, skype, bing and skittle to refer to Black people, Jews, Chinese, and  Muslims, respectively) using SVM classifier based on bag-of-words feature vectors. \citet{elsherief2021latent} introduced a benchmark corpus of 22.5K tweets to study implicit hate speech. The authors presented baseline results over this dataset using Jigsaw Perspective\footnote{https://www.perspectiveapi.com}, SVM, and different variants of BERT \cite{devlin2018bert}.

The use of demographic features such as gender and location in the detection of hate speech is explored by \citet{waseem2016hateful}, and user meta features, e.g., account age, posts per day, number of followers/friends, are used by \citet{ribeiro2017like}.

Computational methods for the detection of hate speech and abusive language range from SVM and logistic regression \cite{davidson2017automated,waseem2016hateful,nobata2016abusive,magu2017detecting}, to neural architectures such as RNNs and CNNs \cite{zhang2016hate, gamback2017using,del2017hate,park2017one}. 
Transformer-based architectures achieved significant improvements, see \cite{mozafari2019bert,aluru2020deep,samghabadi2020aggression,salminen2020developing,qian2021lifelong,kennedy2020contextualizing,arviv2021sa}, among others. 
In an effort to mitigate the need for extensive annotation some works use transformers to generate more samples, e.g., \cite{vidgen2020learning,wullach2020towards,wullach2021fight}. \citet{zhou2021hate} integrate features from external resources to support the model performance.

In order to account for the sometimes elusive and coded language and the unfortunate variety of targeted groups \cite{schmidt2017survey, ross2017measuring}, a set of functional test was suggested by  \citet{rottger2020hatecheck}, allowing an quick evaluation of hate-detection models.

\paragraph{Classification of hate users} Characterizing \emph{accounts} that are instrumental in the propagation of hate and violence is gaining interest from the research community and industry alike, whether in order to better understand the phenomena or in order to suspend major perpetrators instead of removing sporadic content.    
Detection and characterization of hateful Twitter and Gab users was tackled by \citet{ribeiro2018characterizing,mathew2018analyzing,mathew2019spread,arviv2021sa}, among others. An annotated dataset of a few hundreds of Twitter users was released as part of a shared task in CLEF 2021, see \cite{bevendorff2021b} for an overview of the data and the submissions. An annotated dataset of Twitter users using the ambiguous \texttt{\small ((()))} (`echo') symbol was released by \citet{arviv2021sa}.

\paragraph{Hate speech on Parler and Gab} While most prior work focus on the manifestations of hate in the mainstream platforms, a number of works do address alternative platforms such as Gab and Parler. Two annotated Gab datasets were introduced by \citet{kennedy2018gab} and by \citet{qian2019benchmark}. We use these datasets in this work as we compare Parler to Gab. 

Focusing on users, rather than posts, \citet{das2021you} experiment with an array of models for hate users classification. \citet{lima2018inside} aims to understand what users join the platform and what kind of content they share, while \citet{jasser2021welcome} conduct a qualitative analysis studying Gab's platform norms, given the lack of moderation. \citet{gallacher2021hate} explore whether users seek out Gab in order to express hate, or that the toxic attitude is adopted after joining the platform. The spread of hate speech and the diffusion dynamics of the content posted  by hateful and non-hateful Gab users is modeled by  \citet{mathew2019spread} and \citet{mathew2020hate}. 

\ignore{
lima2018inside - characterize Gab, aiming at understanding who are the users who joined it and what kind of content they share in this system

mathew2020hate -  temporal analysis of hate speech on Gab.com

mathew2019spread - the diffusion dynamics of the posts made by hateful and non-hateful users on Gab

gallacher2021hate  explores how hate speech develops, We investigate whether users seek out this platform in order to express hate, or whether instead they develop these opinions over time through a mechanism of
socialisation, as they interact with other users on the platform

das2021you - run a detailed exploration of the hate users classification problem and investigate an array of models ranging from purely textual to graph based to finally semi-supervised techniques using Graph Neural Networks
(GNN) that utilize both textual and graph-based features.

kennedy2020contextualizing - propose a novel regularization based approach in order to increase model sensitivity to the context surrounding group identifiers.

jasser2021welcome - conduct a qualitative analysis of
the Gab. they  find Gab’s technological affordances – including its lack of content moderation, culture of anonymity, microblogging architecture and funding model – have fostered an ideologically eclectic far-right community united by fears of persecution at the hands of ‘Big Tech’.

qian2019benchmark - introduce two fully-labeled large-scale hate speech intervention datasets collected from Gab and Reddit
}

Parler, launched in August 2018 and experiencing its impressive expansion of user base from late in 2020, is only beginning to draw the attention of the research community. Early works analysed the language in Parler in several aspects such as QAnon content \cite{sipka2021comparing}, COVID-19 vaccine \cite{baines2021scamdemic}, and the 2021 Capitol riots \cite{esser2021does}.
The first dataset of Parler messages was introduced by \citet{aliapoulios2021early}, along with a basic statistical analysis of the data, e.g., the number of posts and the number of registered users per month, along with the most popular tokens, bigrams, and hashtags in the data. \citet{ward2021parlez} used a list of predefined keywords (hate terms), assessing the level of hate-speech on the platform.

Our work differs from these works in a number of fundamental aspects. First, we combine textual and social (network) signals in order to detect both hateful posts and hate-promoting accounts. Second, We suggest models that rely on state-of-the-art neural architectures and computational methods, while previous work detects hate speech by matching a fixed set of keywords from a predefined list of hate terms. Furthermore, we provide a thorough analysis of the applicability of different algorithms, trained and fine-tuned on various datasets and tasks. Third, we provide a broader context to our analysis of the proliferation of hate in Parler, as we compare and contrast it to trends observed on Gab. 

\section{Data}
\label{sec:data}
In this section we describe the datasets used for this work -- starting with a general overview of the platforms, then providing a detailed description of the datasets and the annotation procedure. 

\subsection{Parler and Gab Social Platforms}
\label{subsec:parler_gab_social_platfomrs}
\paragraph{Parler} Alluding to the french verb `to speek', Parler was launched on August 2018. The platform brands itself as ``The World's Town Square''  a place in which users can \emph{``Speak freely and express yourself openly, without fear of being ``deplatforme'' for your views.''}\footnote{Parler branding on its landing page (accessed: 1/10/2022).}. 

Parler users post texts (called \emph{parlays}) of up to 1,000 characters. Users can reply to parlays and to previous replies. 
Parler supports a reposting mechanism similar to Twitters retweets (referred to as `echos').  Throughout this paper we refer to echo posts as \emph{reposts}, not to confuse with the ((())) (echo) hate symbol.

Parler's official guidelines\footnote{parler.com/documents/guidelines.pdf (accessed: 1/15/2022)} explicitly allow ``trolling'' and ``not-safe-for-work'' content, include only two ``Principles'' prohibiting ``unlawful acts'', citing ``Obvious examples include: child sexual abuse material, content posted by or on behalf of terrorist organizations, intellectual property theft.'' and spamming. 

By January 2021, 13.25M users have joined Parler and its mobile application was the most downloaded app in Apple's App Store. This growth is attributed to celebrities and political figures promoting the platform (see Section \ref{sec:intro}) and the stricter moderation enforced by Facebook and Twitter, culminating with the suspension of the \texttt{@realDonaldTrump} account from Twitter and Facebook.   

\paragraph{Gab}  Gab, launched on August 2016, was created as an alternative to Twitter and it positioning itself as putting ``people and free speech first'' and welcoming users suspended from other social networks \cite{zannettou2018gab}.
Gab posts (called \emph{gabs}) are limited to 300-characters, and users can repost, quote or reply to  previously created gabs. Gab permits pornographic and obscene content, as long as it is labeled \emph{NSFW} (Not-Safe-For-Work). Previous research finds that Gab is a politically oriented system -- while many users who use the platform are extremists, the majority of users are Caucasians-conservatives-males \cite{lima2018inside}. For more details about gab usage, users and manifestations of hate see references at Section  \ref{sec:related_work}.

\subsection{Parler and Gab Corpora}
\label{parler_and_gab_corpora}
We use the Parler and Gab datasets published by \citet{aliapoulios2021early} and \citet{zannettou2018gab}, respectively. The Parler dataset is unlabeled, therefore annotation is required. We describe the annotation procedure and label statistics in Section \ref{subsec:annotated_data}.

Both datasets include posts and users' meta data, though the Parler dataset is richer, containing more attributes such as registration time and total number of likes.
Each of the datasets is composed of millions of posts and replies, see Table \ref{table:Data_statistics}. The Parler dataset is bigger, containing more posts and more users, however, on average, Gab users post more content per user. We note that there is no temporal overlap between the two datasets. We discuss this point and its impact on the analysis and comparison in Section \ref{sec:discussion}.

\begin{table}[t]
\centering
{
    \begin{tabular}{l@{ }|c@{\quad}c@{\quad}}
      {} & {Parler} & {Gab}\\[2pt]
      \hline\rule{0pt}{12pt}Users       &4.08M       &144.3K \\[2pt]
        Posts	    &20.59M      &7.95M \\[2pt]
        Replies 	&84.55M      &5.92M \\[2pt]
        Reposts     &77.93M      &8.24M \\[2pt]
        Time-Span   &08/2018 -- 01/2021 & 08/2016 -- 01/2018 \\[2pt]
    
    \end{tabular}
   \caption{Datasets Statistics. Replies are comments to main posts. Reposts are equivalent to retweets in Twitter.}
   \label{table:Data_statistics}
}
\end{table}

We use three Gab \emph{annotated} datasets which are all sampled from the unlabeled Gab corpus we use:  (i) The Gab Hate Corpus -- 27.5K Gab posts published by \citet{kennedy2018gab}, (ii) 9.5K Gab posts published by \citet{qian2019benchmark}, and (iii) 5K posts published by \cite{arviv2021sa}. 
In total, we collect a corpus of 42.1K annotated Gab posts. 7.7K (18.4\%) of the posts are tagged as hateful.

\ignore{
    \begin{table}
    \centering
    {\footnotesize
        \begin{tabular}{l@{\quad}|c@{\quad}c@{\quad}c@{\quad}|c@{\quad}c@{\quad}c@{\quad}}
        &
        \multicolumn{3}{c|}{Parler (4.08M Users)} &
        \multicolumn{3}{c}{Gab (144.3K Users)} \\[2pt]
        & Avg. & Med. & Std.& Avg. & Med. & Std.\\
        \hline
        Posts &8.69 &1 &183.33 &66.67 &3 &609.9\\
        Replies &35.26  &3 &2238.74 &75.10 &4 &515.17\\
        Reposts &101.42  &5 &751.62 &116.66 &4 &1095.9\\
        Age     &127.13  &63 &116.97 &260.66 &242 &162.95 \\
        $\overleftarrow{Follow}$ &93.29  &3 &7792.27 &90.86 &12  &468.19 \\
        $\overrightarrow{Follow}$ &75.14  &12 &918.88 &99.42 &5  &1104.31 \\
        \end{tabular}
    }
        \caption{User level measures. Age is measured in days. Med. is the median. $\overleftarrow{Follow}$ and $\overrightarrow{Follow}$ are the measures for the followers and followees respectively.}
       \label{table:statistics_users_perspective}
    \end{table}
}

\subsection{Parler Data Annotation}
\label{subsec:annotated_data}
Hate speech takes different forms in different social platforms \cite{wiegand2019detection} and across time \cite{florio2020time}. It is often implicit \cite{elsherief2021latent}, targeting a variety of groups. Consequently, transfer learning remains a challenge for hate-speech detection, and annotated Parler data is needed in order to achieve accurate classification. This challenges and the significant improvements in performance achieved by proper fine-tuning are demonstrated through extensive experimentation, see Section \ref{subsec:post_level_classification}. In the remainder of this section we describe the annotation procedure and the annotated dataset we use. 

The annotation task was designed as follows: 10K posts were sampled from the full Parler corpus. All posts met the following criteria: (i) Primary language is English; (ii)  A post should be at least 10 characters long; (iii) The post does not contain a URL; and (iv) The post is neither a repost nor a comment.

The 10K annotated posts \emph{were not} randomly selected from the Parler corpus. A random selection of posts would have led to an extremely imbalanced dataset as most of the posts do not contain hate speech. Hence, we opt to stratified sampling. This sampling process relies on an approximation of the likelihood of each post to include hateful content. We used a pretrained hate speech prediction model to approximate this likelihood.

Annotation was done by 112 student (more than half of them are graduate students), who were provided detailed guidelines and training involving the various types of hate speech, the elusiveness of hate expressions using coded language, how to detect it, and a number of examples of different types. Each of the annotators was prompted with a list of 300 posts and had to assign each with a Lickert score ranging from 1 (not hate) to 5 (extreme or explicit hate). We provided annotators only with the textual content of the post. Each of the 10K posts was annotated by three annotators.
Annotators presented a satisfying agreement level of 72\% and a Cohen's Kappa of 0.44. Labels of posts with a low agreement level\footnote{We define low agreement as posts labeled with least three unique values or if the difference between annotations was higher than 2.} were ignored ($\sim$7\% of the annotated posts). We define a post as hateful (non-hateful) if its average score is higher (lower) than three. We omit posts with an average score of exactly three. Accordingly, 3224 of the 10K posts (32.8\%) were labeled as hateful and 6053 (59.8\%) as non-hateful.

We make this annotated corpus available in the project's repository\footnote{https://github.com/NasLabBgu/parler-hate-speech} -- the first public annotated corpus of Parler.

\begin{figure*}
    \centering
    \small
    \includegraphics[scale=0.18]{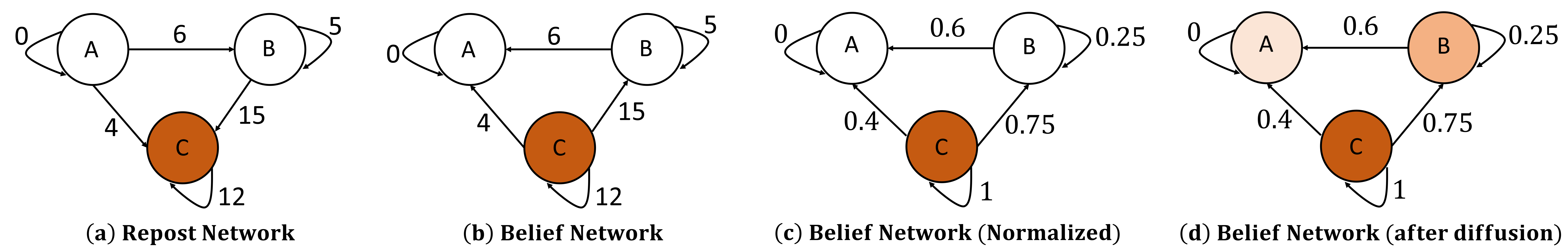}
    \caption{\small An illustration of the diffusion model over three nodes. Self loops represent the total number of posts per node. In step (a) we build the repost network and assign each node with an initial belief -- seed hate mongers with a value of one and others with a value of zero. In steps (b) and (c) we convert the network to a belief network -- reversing the edges' direction and normalizing their weight. In step (d) we run the diffusion process and get a belief score per node, which is indicated in the graph by the darkness of each node.}
    \label{fig:diffusion}
\end{figure*}

\section{Methods}
\label{sec:methods}
In this work we are interested in the detection of hate, both on the post level and the account level. Our interest in the post level classification is twofold. Given an accurate classifier, we can: (a) Approximate the hate degree in different aggregation levels -- e.g., over all social network, and per user, and  (b) Use the post-level predictions to support training a user level classifier. A review of the various post level classifiers is provided in Section \ref{subsec:post_level_classification} and our modifications to a diffusion-based model for user classification are presented in Section \ref{subsec:user_level_classification}. Ethical considerations related to user classification are discussed at the end of Section \ref{sec:discussion}.

\subsection{Post Level Classification Models}
\label{subsec:post_level_classification}

We fine-tune the DistilBERT \cite{Sanh2019DistilBERTAD} transformer on each of the datasets, obtaining two fine-tuned models (referred to as Our-FT BERT). We compare the performance of the models on the respective datasets against four competitive models:
\begin{enumerate} 
    \item {\bf Jigsaw Perspective}: A widely used commercial model to detect hate and toxic content, developed by Google. Jigsaw was found to perform well in an array of tasks related to hate-speech detection \cite{rottger2020hatecheck}. Jigsaw implementation is not public and the service is provided as a black-box through an online API\footnote{https://www.perspectiveapi.com}.
    \item {\bf deHateBERT} \cite{aluru2020deep}: An adaptation of the BERT Transformer for hate-speech detection -- the pretrained transformer was fine-tuned on a corpus of 96.3K text snippets from Twitter and from the white supremacist forum Stormfront.org. The authors indicate that 15.01K (15.6\%) training samples were labeled as hate-speech.
    \item  {\bf Twitter-roBERTa} \cite{barbieri2020tweeteval}: This model uses the RoBERTa \cite{liu2019roberta} architecture, specifically fine-tuned on the task of hate-speech detection of micro-messages. The authors used a corpus of 13K tweets, 5.2K (40\%) of them are labeled as hate speech.
    \item {\bf HateBase} \cite{tuckwood2017hatebase}: HateBase is a multilanguage vocabulary of hate terms that is maintained on order to assist in content moderation and research. We use 68 explicit hate terms that were used in prior works \citet{mathew2018analyzing,mathew2019spread}. These terms were mainly selected from HateBase's English lexicon and is composed only of \emph{explicit} hate terms like `kike' (slur targeting Jews, see post \#2 in Table \ref{table:posts_sample}), `paki' (slur against Muslims, especially with Pakistani roots), and `cunt' (see post \#1 in Table \ref{table:posts_sample}). 
    
\end{enumerate}

\subsection{User Level Classification}
\label{subsec:user_level_classification}

Ideally, an account should be classified as a hate account based on the content it posts (or likes). However, this seemingly straight forward approach is severely limited by ambiguity, vagueness, dog-whistling, and emerging idioms and racial slurs. For example, defining a threshold of $k$ hateful posts is still not well defined. How explicit these $k$ posts should be? would $2k$ less explicit posts make the cut? is one post enough to declare a user a hate-monger? Moreover, defining a threshold does not account for networked aspect of the data and the fact that ``birds of a feather flock together'' \cite{himelboim2013birds}.

In order to leverage the network structure, we view each platform as a social network with users as nodes and \emph{reposts}  as directed edges. 
Edges are weighted to reflect levels of engagement, as illustrated in Figure \ref{fig:diffusion}(a): a directed edge $(A, B)$ with a weight of 6 indicates that user $A$ reposted 6 posts originally posted by user $B$.

We build on the diffusion-based approached for the detection of hate mongers, proposed by \citet{mathew2019spread}, modifying it in order to achieve a more accurate classification. The basic diffusion-based classification is achieved in two stages: (a) Identifying a \emph{seed} group of hate mongers. (b) Applying a diffusion model over the social network. We use the DeGroot’s hate diffusion model \cite{golub2010naive} which outputs an estimated belief value (i.e., ``hate'') per user, over the [0,1] range. A toy example of the diffusion process is illustrated in Figure \ref{fig:diffusion}. In our experiments we set the number of diffusion iterations to three. One clear advantage of this approach over fully supervised methods is that it does not require a large dataset annotated on the user level.

\paragraph{Modified Diffusion Model} We modified the diffusion model used by \citet{ribeiro2018characterizing} and \citet{mathew2019spread} in two ways: (i) \emph{Seed definition}. Instead of taking a lexical approach in order to identify users posting more than $k$ hateful posts,  we use our fine-tuned Transformers. We argue that fine-tuning the classifiers for each social network significantly improves the classification on the post level (as demonstrated in Section \ref{subsec:results_post_level}), and ultimately, improves the performance of the diffusion model; and (ii) \emph{Hateful users definition}. In the original diffusion process, hate (as well as ``not-hate'') labels are diffused through the network. This way, seed hate mongers may end with a low belief (hate) score, which in turn propagates to their neighbours. However, seed users were chosen due to the fact that they post a significant number of undoubtedly hateful posts. Fixing the hate score of these users results in a more accurate labeling of the accounts in the network.

\section{Classification Results}
\label{sec:results}
\subsection{Post Level Results}
\label{subsec:results_post_level}
We use the annotated corpora (see Section \ref{subsec:annotated_data}) to fine-tune the pretrained Transformer  on each social platform, splitting the labeled data to train (60\%), validation (20\%), and test (20\%) sets.

The precision-recall curves of the Parler and Gab models are presented in Figure \ref{fig:pr_curves}. Our fine-tuned models  significantly outperforms the other models in both datasets. 
We wish to point out that while the popular keyword base approach (HateBase) achieves a high precision and a moderate recall on the Gab data, outperforming all Transformer models except the platform fine-tuned ones, it collapses in both measures on the newer Parler dataset. These results revalidate the limitations of lexical approaches, and of neural methods that are not fine-tuned for the specific dataset (even though they were fine-tuned for a similar task  -- hate speech detection in another microblogging platform). 

\begin{figure}\small
    \centering
    \begin{minipage}{0.22\textwidth}
	    \centering
	    \begin{tikzpicture}[scale=0.49]
          \begin{axis}[
              xlabel={Recall},
              ylabel={Precision},
              ytick={0.0, 0.2, 0.4, 0.6, 0.8, 1.0},
              yticklabel style={/pgf/number format/fixed},
              xtick={0.0, 0.2, 0.4, 0.6, 0.8, 1.0},
              label style={font=\large},
              legend style={font=\scriptsize}
            ]
              \node[shape=diamond, fill=orange, scale=0.4] at (axis cs:0.02, 0.71) {};
      
              \addplot[mark=o, color=blue, line width=1.0pt, mark size=0.0pt] table [x=recall, y=precision, col sep=comma] {precision_recall_curves/parler/FT_BERT.csv};
              \addplot[mark=o, color=black, line width=1.0pt, mark size=0.0pt, dashed] table [x=recall, y=precision, col sep=comma] {precision_recall_curves/parler/deHateBERT.csv};
              \addplot[mark=x, color=violet, line width=1.0pt, mark size=0.0pt, dashdotted] table [x=recall, y=precision, col sep=comma] {precision_recall_curves/parler/perspective.csv};
              \addplot[mark=o, color=red, line width=1.0pt, mark size=0.0pt, dotted] table [x=recall, y=precision, col sep=comma] {precision_recall_curves/parler/Twitter_roBERTa.csv};
              
              \legend{Our-FT-BERT, deHateBERT, Jigsaw, Twitter-roBERTa}
            \end{axis}
        \end{tikzpicture}
        \caption*{\small Parler Classifiers}
    \end{minipage}
    \begin{minipage}{0.22\textwidth}
        \centering
        \begin{tikzpicture}[scale=0.49]
            \begin{axis}[
              xlabel={Recall},
              ylabel={Precision},
              ytick={0.0, 0.2, 0.4, 0.6, 0.8, 1.0},
              yticklabel style={/pgf/number format/fixed},
              xtick={0.0, 0.2, 0.4, 0.6, 0.8, 1.0},
              label style={font=\large},
              legend style={font=\scriptsize}
          ]
          \node[shape=diamond, fill=orange, scale=0.4] at (axis cs:0.33, 0.91) {};
      
          \addplot[mark=o, color=blue, line width=1.0pt, mark size=0.0pt] table [x=recall, y=precision, col sep=comma] {precision_recall_curves/gab/FT_DistilBERT.csv};
          \addplot[mark=o, color=black, line width=1.0pt, mark size=0.0pt, dashed] table [x=recall, y=precision, col sep=comma] {precision_recall_curves/gab/RM_deHateBERT.csv};
          \addplot[mark=x, color=violet, line width=1.0pt, mark size=0.0pt, dashdotted] table [x=recall, y=precision, col sep=comma] {precision_recall_curves/gab/perspective.csv};
          \addplot[mark=o, color=red, line width=1.0pt, mark size=0.0pt, dotted] table [x=recall, y=precision, col sep=comma] {precision_recall_curves/gab/RM_Twitter_roBERTa.csv};
          
          \legend{Our-FT-BERT, deHateBERT, Jigsaw, Twitter-roBERTa}
        \end{axis}
      \end{tikzpicture}
      \caption*{\small Gab Classifiers}
    \end{minipage}

  \caption[Precision-Recall (PR) curves -- post level]{\small Precision-Recall (PR) curves -- post level. Results are over the test set. FT-BERT stands for Fine Tuned BERT. The orange diamond (\textcolor{orange}{$\Diamondblack$}) marks the PR performance of the lexical-based approach (HateBase). Unlike the other four methods, this approach cannot be controlled by a threshold parameter, hence only a single PR value is available.}
\label{fig:pr_curves}
\end{figure}

\subsection{User Level Results}
\label{subsec:results_user_level}
As described in Section \ref{subsec:user_level_classification}, in order to classify accounts we use a diffusion model. The diffusion process is seeded with a set of hateful accounts. The choice of seed accounts involves the following steps: (i) After establishing the accuracy of the fine-tuned models (Section \ref{subsec:results_post_level}) we use these models to label \emph{all} the posts in the respective dataests. (ii) Opting for a conservative assignment of seed users, we consider only posts with hate score (likelihood) over 0.95 (0.9) in the Parler (Gab) dataset to be hateful. Finally, (iii) Users posting 10 or more hateful posts are labeled as seed accounts. We take the conservative approach in steps (ii) and (iii) in order to control the often noisy diffusion process.  

\begin{figure}
\centering
\begin{tikzpicture}
  \centering
  \begin{axis}[
        ybar=4pt, axis on top,
        bar width=20pt,
        height=5.8cm, width=7.5cm,
        bar width=0.65cm,
        ymin=0, ymax=60,
        axis x line*=bottom,
        y axis line style={draw=none},
        ytick style={draw=none},
        ytick=\empty,
        tickwidth=5pt,
        enlarge x limits=0.6, 
        legend style={
            at={(0.8,1.09)},
            legend columns=-1,
            /tikz/every even column/.append style={column sep=0.4cm}},
        ylabel near ticks, yticklabel pos=left,
        ylabel={Active Users (\%)},
        symbolic x coords={Praler,Gab},
       xtick=data,
       nodes near coords={
        \pgfmathprintnumber[precision=1]{\pgfplotspointmeta}
       },
       nodes near coords style={}
    ]
    \addplot [draw=none, style={fill=myred, postaction= {}}] coordinates {
      (Praler, 16.1)
      (Gab, 10) };
   \addplot [draw=none, style={fill=mypurple, postaction= {pattern=north east lines}}] coordinates {
      (Praler, 42.4)
      (Gab, 41.7) };
    
    \addplot [draw=none, style={fill=myblue, postaction= {pattern=grid}}] coordinates {
      (Praler, 41.5)
      (Gab, 48.3) };

    \legend{$HM$, $\widetilde{HM}$, $N$}
    
  \end{axis}
  \end{tikzpicture}
  
  \caption{Distribution of the \emph{active} users in Parler and Gab}
  \label{fig:hate_mongers_distribution}
\end{figure}
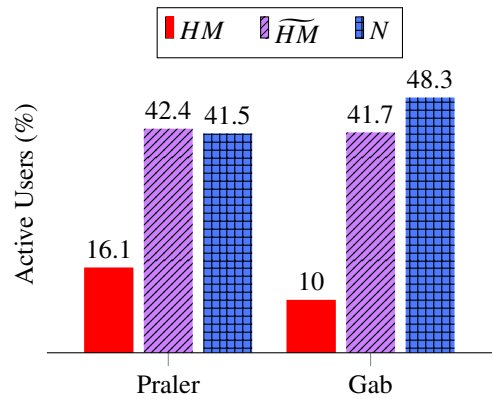

Simulating the modified diffusion process described in Section \ref{subsec:user_level_classification} we obtain a hate score per \emph{user}. For analysis purposes we divide users to three distinct groups -- hate mongers (denoted $HM$), composed of the users making the top quartile of hate scores; normal users (denoted $N$) making the bottom quartile; the rest of the users (denoted $\widetilde{HM}$) suspected as ``flirting'' with hate mongers and hate dissemination. 
Users with a low level of activity (less than five posts or users joining the network in the last 60 days) were not considered\footnote{87.1\% (63.4\%) of the users in Parler (Gab).}. The distribution of \emph{active} users by type is presented in Fig \ref{fig:hate_mongers_distribution}.

\paragraph{Evaluation of the diffusion model} A  user-level annotated dataset of 798 Gab users was shared by \citet{das2021you}. We use this dataset to validate the performance of the diffusion models -- both the standard and our modified models (see Section \ref{subsec:user_level_classification}). We find our modified model to outperform the standard models, achieving precision/recall/F1-scores of 0.9/0.54/0.678, comparing to of 0.95/0.34/0.5. Therefore, results and analysis in the remainder of the paper are based on the modified diffusion model.

\ignore{
    \begin{table}
    \centering
    {\footnotesize
        \begin{tabular}{l@{\quad}|c@{\quad}c@{\quad}|c@{\quad}c@{\quad}}
        &
        \multicolumn{2}{c|}{Parler} &
        \multicolumn{2}{c}{Gab} \\[2pt]Population& Number & \% & Number & \%\\
        \hline
        \emph{$HM$}             &84.7K &16.1 &5.3K &10\\
        \emph{$\widetilde{HM}$} &222.8K &42.4 &22K &41.7\\
        \emph{N}                &217.8K &41.5 &25.5K &48.3 \\
        \hline
        Total &525.3K &100 &52.8K &100\\
        \end{tabular}
    }
        \caption{Diffusion model results. Each user is assigned to one of the three populations. Users with a low activity level or with a short experience in the platform are omitted.}
       \label{table:user_type_distribution}
    \end{table}
}

\section{Hate Analysis}
\label{sec:analysis}

In this section we provide a comprehensive analysis of the propensity for hate speech on Parler and Gab

\subsection{Hate on the Post Level}
\label{subsec:post_level_analysis}
Taking our conservative approach, we find that hate posts are more frequent in Parler (3.29\%) than in Gab (2.13\%). However, we find that 13.95\% of Parler users share at least one hateful post -- significantly lower number compared to Gab (18.58\%). We find that 65.5\% of the hate content in Parler is posted as a reply to other parlays. This reflects a significant over-representation of replies compared with full corpus distribution (46.2\% of posts are replies, see Table \ref{table:Data_statistics}). Similarly, 38.9\% of the hate content on Gab are replies.

\subsection{Hate on the User Level}
\label{subsec:user_level_analysis}
We provide an analysis of the characteristics of the $HM$, $\widetilde{HM}$ and $N$ accounts on an array of attributes, ranging from activity levels to centrality, sentiment and the emotions they convey. 

\paragraph{Activity level} Activity levels are compared via four features -- number of posts, replies, reposts, and users' age (measured in days).

\begin{figure}
\normalsize
    \centering
    \subfloat[{Parler\label{subfig:parler_meta_features}}]{{\includegraphics[scale=0.042]{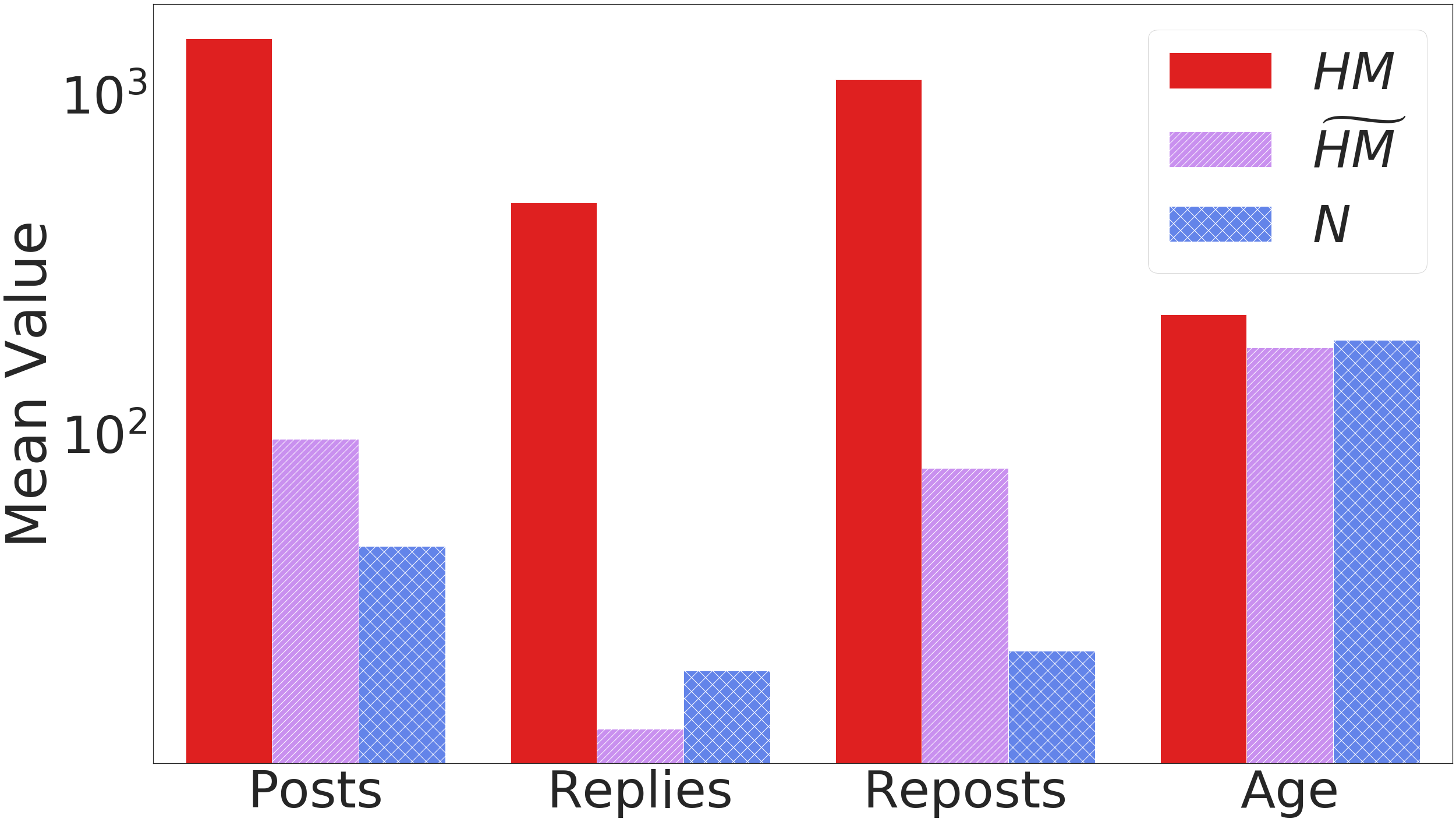}}}
    \quad
    \subfloat[Gab\label{subfig:followers_gab}]{{\includegraphics[scale=0.042]{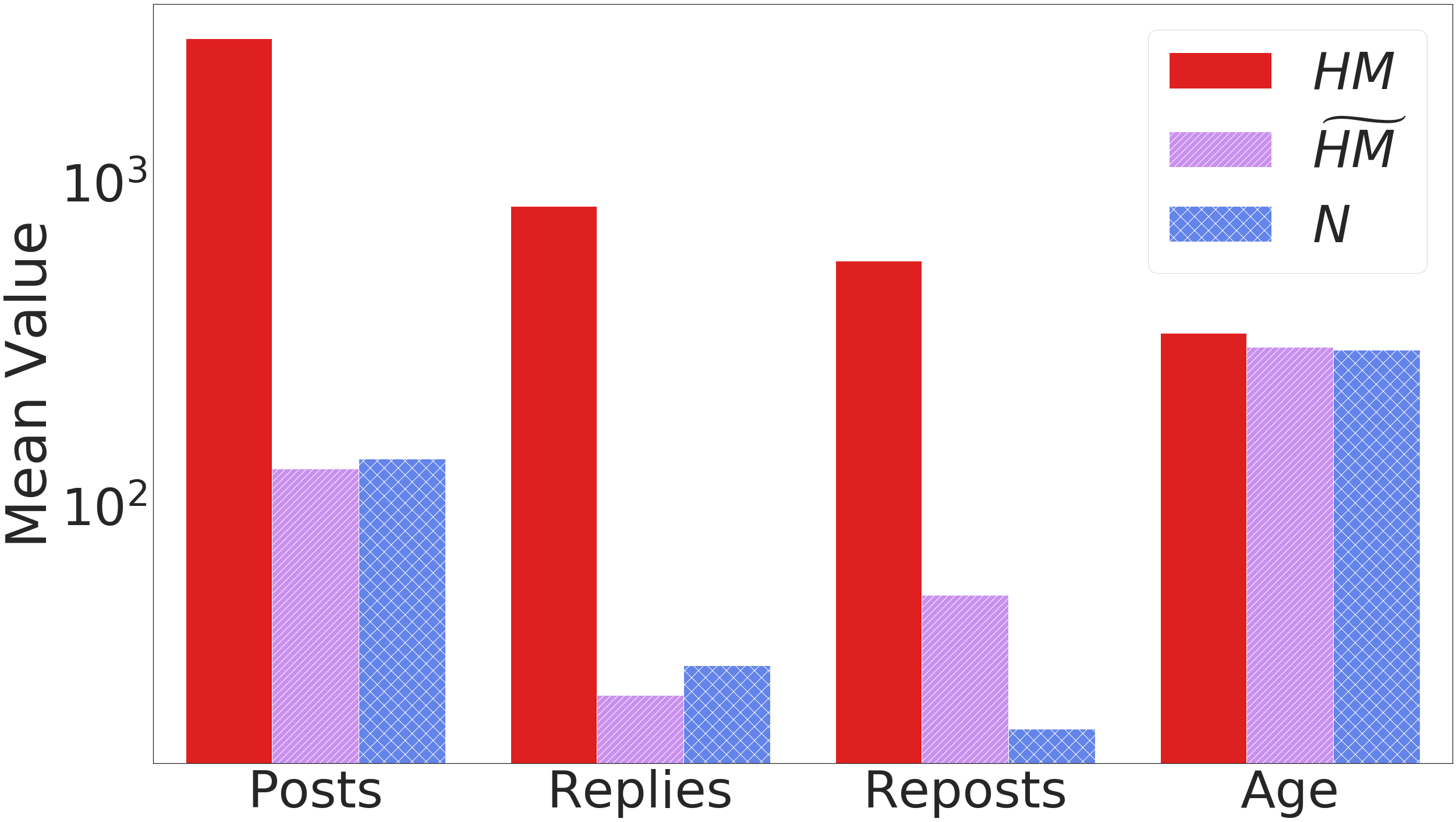}}}
   \caption{\small Activity measures per user group. Numbers are averaged per measure and group. We use a log-scale over the y-axis.}
    \label{fig:activity_measures}
\end{figure}

$HM$ are the most active user group in both platforms across all activity types (see Figure \ref{fig:activity_measures}). We find that the $\widetilde{HM}$ users have similar characteristics in both platforms -- they share less content than the $HM$ users, repost more content than the $N$ group, and their tendency to reply is lower compared to the $N$ users.

Interestingly, although the $HM$ make only 16.1\% (10\%) of the active users in Parler (Gab) -- they generate a disproportional number of posts:  30.6\% (59.45\%) of the posts in Parler (Gab). The same holds for replies -- the $HM$ users post 36.68\% (75.57\%) of the replies in Parler (Gab). When aggregating all activity types (post/reply/repost) -- the $HM$ users generate 41.23\% (71.38\%) of the content in Parler (Gab).

User \emph{Age} (days from account creation to the most recent post in the data), is an exceptional feature. We observed only insignificant differences between the three user groups. This observation holds for both platforms. However, collapsing the groups -- we do find a significant difference between the two platforms. Gab users are ``older'' with an average age of 323.9 compared to 189.6 of the Parler users.
We hypothesize that the difference is a result of the way both platform evolve over time, given the unfolding of events driving users to these platforms (see Sections \ref{sec:intro} and \ref{subsec:parler_gab_social_platfomrs}).

\begin{figure}
\normalsize
    \centering
    \subfloat[Parler Followers\label{subfig:followers_parler}]{{\includegraphics[scale=0.115]{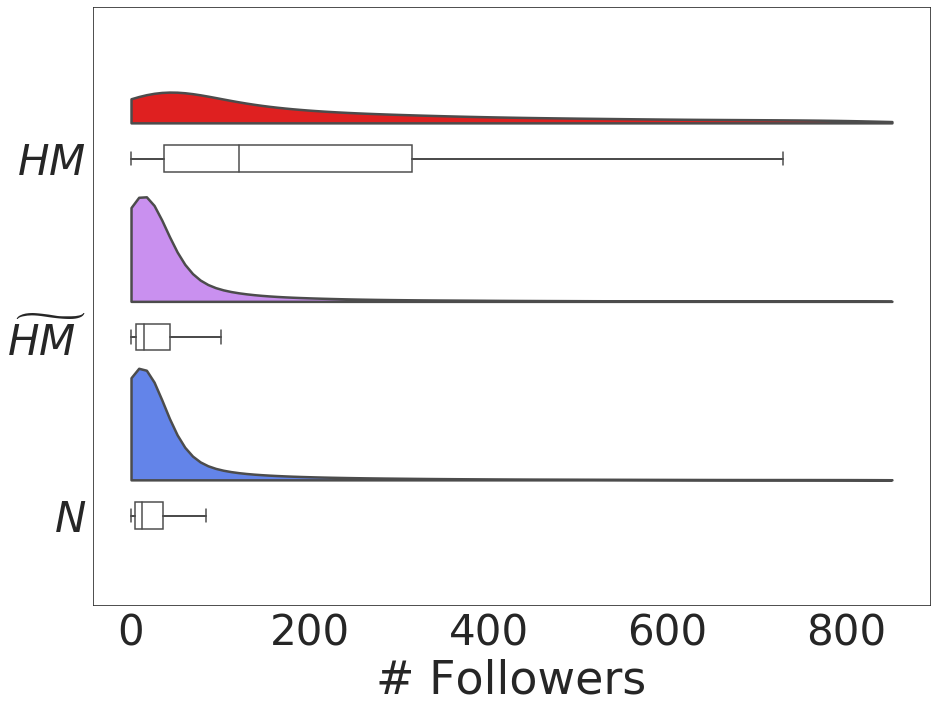}}}
    \quad
    \subfloat[Gab Followers\label{subfig:followers_gab}]{{\includegraphics[scale=0.115]{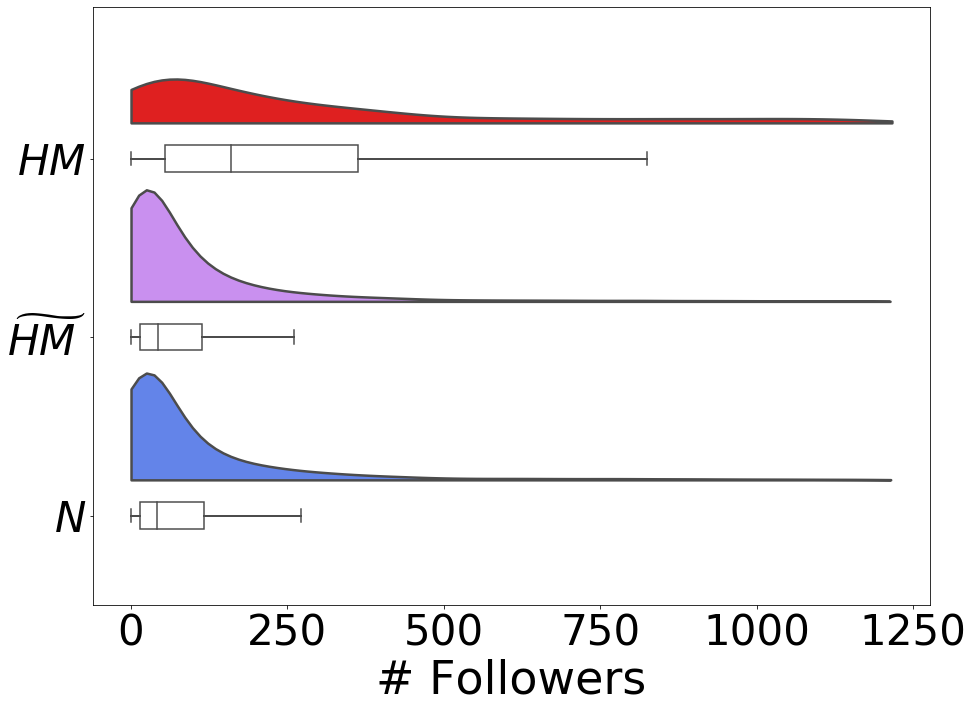}}}
   \caption{\small Followers distributions. Followers are the number of accounts that follow a user. The extreme percentiles (2.5\%) of the data are omitted for visualization purposes. Rectangles below each distribution are the $\pm$ standard division around the average; The vertical line in each rectangle represent the median.}
    \label{fig:followers}
\end{figure}

\paragraph{Popularity and Engagement}
We quantify the popularity level of users based on the number of \emph{followers} they have. Figure \ref{fig:followers} presents numbers for both platforms. On both platforms hate mongers ($HM$) are significantly more popular compared to all other user groups. In Parler, the median number of followers is 121 compared to 15 and 12 of $\widetilde{HM}$ and $N$, respectively. The same holds for Gab -- a median value of 160 for $HM$ users compared to 43 and 41 of the other two user groups. Interestingly, although Parler is a much larger social platform (mainly in terms of registered users, see  Section \ref{sec:data} and Table \ref{table:Data_statistics}) we do not see a significant higher number of followers in Parler. Moreover, when calculating the number of followers over the whole population, the median in Gab is three times higher -- 48 vs. 16.

\ignore{
    \begin{figure*}
    \normalsize
        \centering
        \subfloat[Parler Followees\label{subfig:followees_parler}]{{\includegraphics[scale=0.27]{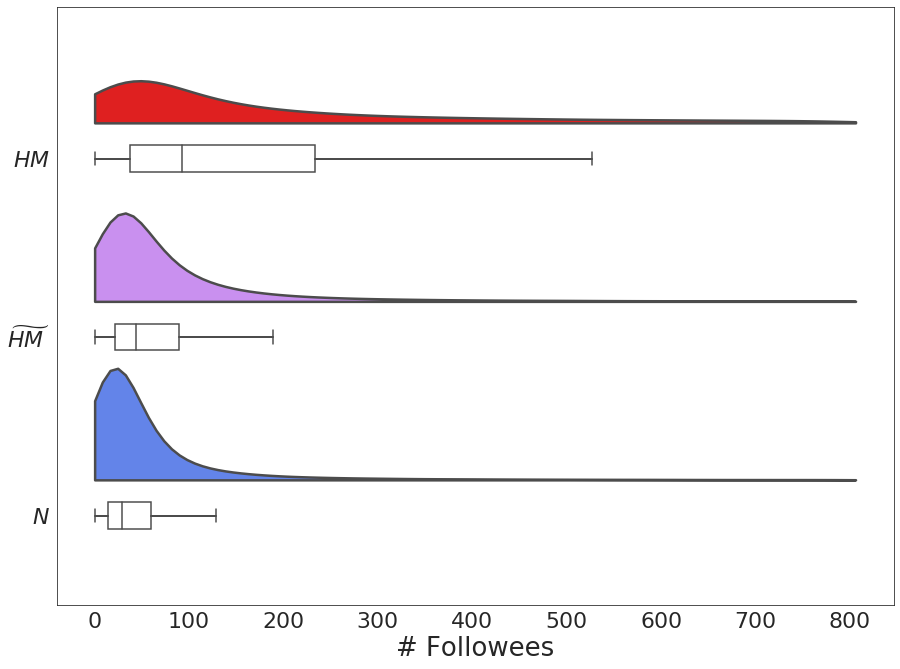}}}
        \quad
        \subfloat[Gab Followees\label{subfig:followees_gab}]{{\includegraphics[scale=0.27]{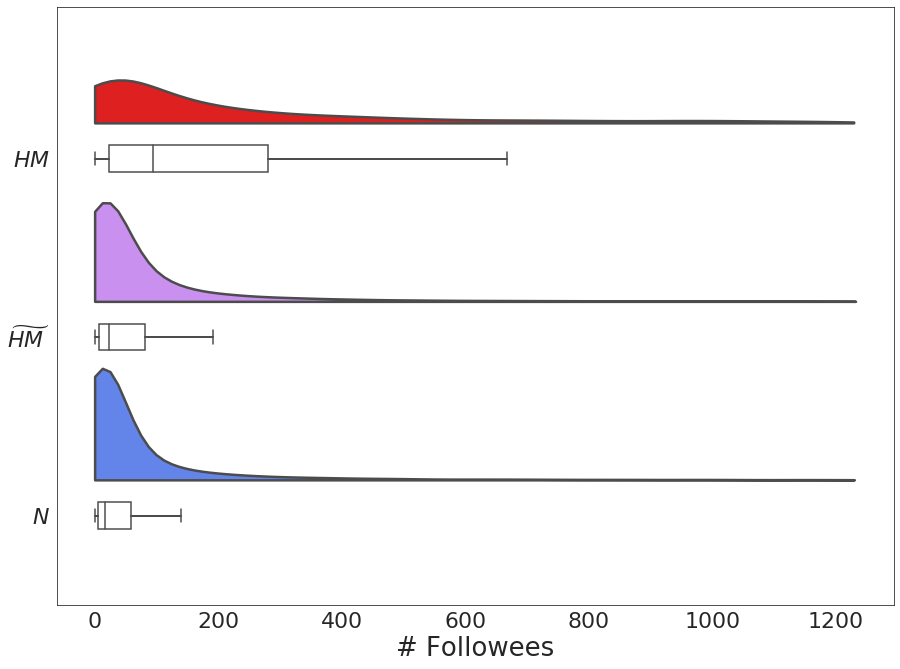}}}
       \caption{\normalsize Followees distributions - Parler (left) and Gab (right). Followees are the number of accounts that a user follows. The extreme percentiles (2.5\%) of the data are omitted for visualization purposes. Rectangles below each distribution are the $\pm$ standard division around the average; The vertical line in each rectangle represent the median.}
        \label{fig:followees}
    \end{figure*}
}

Engagement level is measured by the number of \emph{followees} each account has (the number of accounts a user follows). We find that hate mongers are highly engaged in both platforms compared to other user groups. In Parler, the median followees number of $HM$ users is 106 -- significantly higher than 46 and 36 median values of the $\widetilde{HM}$ and $N$ users, respectively. A similar pattern holds for Gab. 

\ignore{
    \begin{table*}
    \centering
    {\small
        \begin{tabular}{l@{\quad}|c@{\quad}c@{\quad}c@{\quad}c@{\quad}|c@{\quad}c@{\quad}c@{\quad}c@{\quad}|c@{\quad}c@{\quad}c@{\quad}c@{\quad}}
        \multicolumn{1}{c|}{} &
        \multicolumn{4}{c|}{$HM$} &
        \multicolumn{4}{c|}{$\widetilde{HM}$} &
        \multicolumn{4}{c}{$N$} \\[2pt]
        & Avg & Median & $q^{.9}$ & Std.& Avg & Median & $q^{.9}$ & Std.& Avg & Median & $q^{.9}$ & Std.\\
        \hline
        Biography Exists &0.596 &- &- &- &0.338 &-  &- &- &0.358 &-  &-  &-\\
        Biography Length &128.6  &134 &198 &71.5 &99.4 &90 &191 &62.1 &94.6 &84 &189 &61.8\\
        \end{tabular}
    }
        \caption{Biography measures. We present statistics for Parler only, as this information is unavailable for Gab. `Biography Exists' is an indicator variable whether a biography string was written by the user (i.e., non mandatory). `Biography Length' is the text length, in case it exists. $q^{.9}$ is the 90\% percentile of the distribution.}
       \label{table:bio_measures}
    \end{table*}
}

\paragraph{Account's self description} Analogue to the account's description in Twitter, Parler users can provide a short descriptive/biographical text to appear next to the user's avatar. For example, the biography that is associated with a specific 
Parler user is: \emph{``Conservative banned by mainstream social media outlets for calling the leftists out for what they really are! Been awake for YEARS! \#trump2020 \#maga''}.

We use this content to further assess users commitment to the platform, assuming more engaged users are, the more likely they add the description to their profile. 
We find that while only 35.8\% of the $N$ users use the biography field, 59.6\% of the $HM$ users provide the description in their profile. 
We also find that the average (median) biographical text length of $HM$ users is 128.6 (134) -- considerably longer, compared to $\widetilde{HM}$ and $N$ users who included the description in their profile, with an average (median) of 99.4 (90) and 94.6 (84), respectively.

\paragraph{Social Structure}
We further analyze the differences between Parler and Gab platforms over the different user groups from a social network analysis (SNA) perspective, based on the reposts network. 
Table \ref{table:structural_features} provides an overview of a number of centrality measures. 
The $HM$ users have a significantly higher values in all measures in both platforms. Interestingly, the full  order between the different user groups is kept only for the `betweeness' centrality, while other centrality measures a less stable comparing the $\widetilde{HM}$ and $N$ groups.

\begin{figure}
\normalsize
    \centering
    \subfloat[Parler\label{subfig:degree_distrib_parler}]{{\includegraphics[scale=0.14]{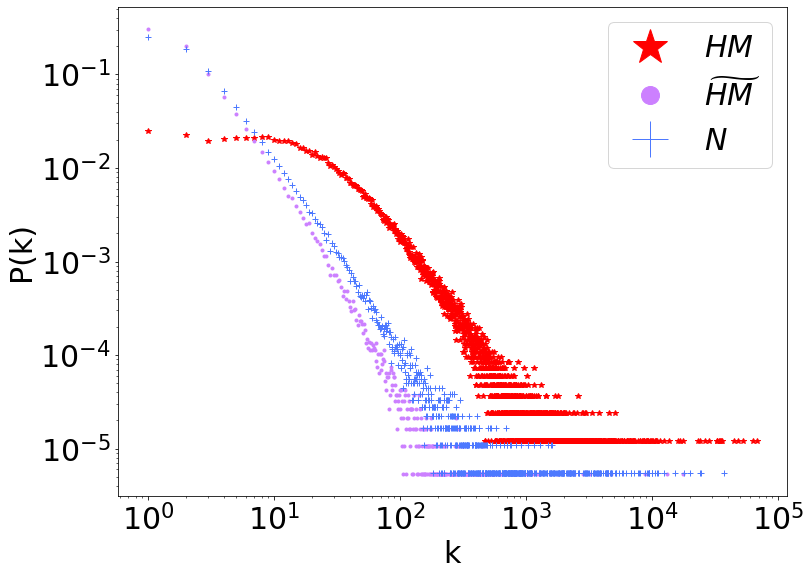}}}
    \quad
    \subfloat[Gab\label{subfig:degree_distrib_gab}]{{\includegraphics[scale=0.14]{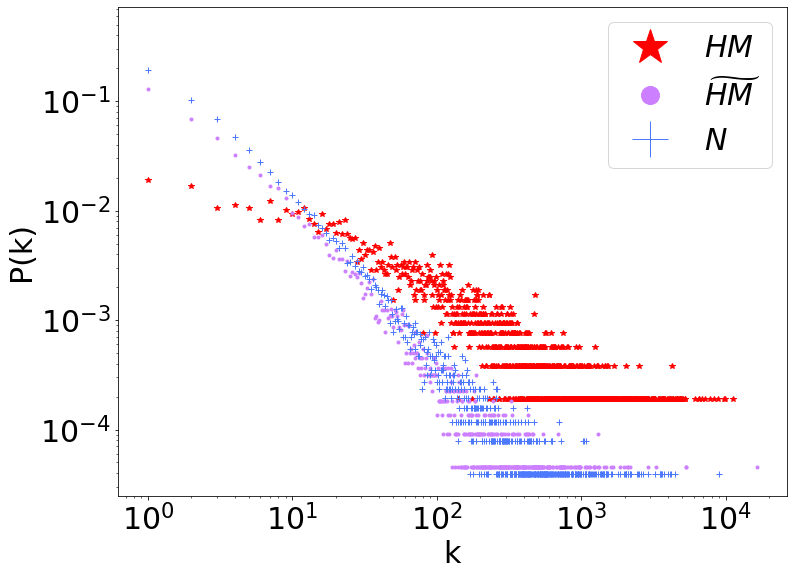}}}
   \caption{\small Social networks degree distribution. We present the in-degree distributions. Network is based on reposts. p(k) (y-axis) is the probability value per a each node's degree (x-axis). We use a log-scale over both the axis.}
    \label{fig:degree_distrib}
\end{figure}

Analysing the degree distribution of users provides an interesting difference between the platforms. In line with the numbers in Table \ref{table:structural_features}, $HM$ users have the most distinctive distribution in both Parler and Gab. However, while the $\widetilde{HM}$ and the $N$ group distributions are inseparable in Gab, the Parler user groups have distinct distributions (see Figure \ref{fig:degree_distrib}). These distributions highlight the distinctiveness of the location of $HM$ users in the network, as well the role of the $\widetilde{HM}$ compared to $N$ users.


\begin{table}
\centering
\footnotesize
{
    \begin{tabular}{c@{\quad}|l@{\quad}|c@{\quad}c@{\quad}c@{\quad}}
    & & $HM$ & $\widetilde{HM}$ & $N$ \\[2pt]
    \hline
    \multirow{3}{*}{\rotatebox{270}{Parler}}&
    ID Centrality&          \num{3.26e-5}   &\num{1.43e-6}   &\num{3.64e-6}\\
    &OD Centrality&         \num{4.01e-5}   &\num{1.97e-6}   &\num{2.31e-6}\\
    &Betweenness&           \num{3.43e-6}   &\num{1.61e-7}   &\num{1.1e-7}\\
    &PageRank&              \num{1.11e-6}   &\num{2.47e-7}  &\num{4.74e-7}\\
    \hline
    \multirow{3}{*}{\rotatebox{270}{Gab}}&
    ID Centrality&          \num{3.35e-3}   &\num{1.18e-4}  &\num{2.81e-4}\\
    &OD Centrality&         \num{3.35e-3}   &\num{4.06e-4}  &\num{9.92e-5}\\
    &Betweenness&           \num{1.43e-4}   &\num{6.11e-6}  &\num{4.54e-6}\\
    &PageRank&              \num{8.85e-5}   &\num{7.17e-6}  &\num{9.68e-6}\\
    \end{tabular}
}
    \caption{\small Structural features. Values are averaged over all users in each user group. `ID' and `OD' are the in-degree and out-degree respectively.}
   \label{table:structural_features}
\end{table}

\paragraph{Linguistic features} We compare the sentiment expressed and the emotions conveyed by different user groups. We use pretrained BERT models for both the sentiment\footnote{https://huggingface.co/nlptown/bert-base-multilingual-uncased-sentiment} and emotion\footnote{https://huggingface.co/bhadresh-savani/distilbert-base-uncased-emotion} predictions. Results are presented in Table \ref{table:sentiment_objectivity}. Looking at the Parler users, we find a small though significant (p-value $< 10^{-3}$) tendency of $HM$ to express a more negative sentiment.  The same holds for Gab, although the sentiment expressed by $\widetilde{HM}$ is closer to the sentiment of the $HM$ users, rather to that of the $N$ users.
Aggregating the emotion predictions, we find that $HM$ users tend to convey more \emph{Anger} and \emph{Sadness} than the other groups. This observation holds for both Parler and Gab, although \emph{Anger} is more prominent.



\begin{table}[h!]
\centering
{\small
    \begin{tabular}{c@{\quad}c@{\quad}|c@{\quad}c@{\quad}c@{\quad}c@{\quad}||c@{\quad}}
    & & Anger & Joy & Sad & Fear & Sentiment\\[2pt]
    \hline
    \multirow{3}{*}{\rotatebox{270}{Parler}}&
    $HM$                &48\% &37.9\% &7.4\% &5.1\% &2.63 \\
    &$\widetilde{HM}$   &41.9\% &44.3\% &6.7\% &5.3\%&2.84 \\
    &$N$                &33.6\% &55.7\% &5\% &4.3\%&2.84\\
    \hline
    \multirow{3}{*}{\rotatebox{270}{Gab}}&
    $HM$                &40.0\% &44.5\% &7.2\% &6.3\%&2.55 \\
    &$\widetilde{HM}$   &35.9\% &49.7\% &5.9\% &7.1\%&2.56 \\
    &$N$                &35.5\% &51.1\% &6.0\% &5.7\%&2.67\\
    \end{tabular}
}
    \caption{\small Emotions and sentiment analysis. The four leftmost columns are the distribution of emotions per user group while the rightmost column is the median sentiment score. The sentiment spans over [1,5] (i.e., 5 is the highest score). We omit Love and Surprise emotions since their proportion in all groups is negligible.}
   \label{table:sentiment_objectivity}
\end{table}

\section{Discussion}
\label{sec:discussion}
\paragraph{Seed hate mongers} One design choice critical to this work, affecting the user-level analysis, is the way we define \emph{seed hate mongers} (see Section \ref{subsec:user_level_classification}). Previous works used lexicon based solutions. We decided to use our post level classification model which significantly outperformed other alternatives. However, both solutions rely on counting the number of hate posts per user.
This binary definition lacks the sensitivity to mark hate mongers in a more nuanced way.  
Alternative methods to mark seed hate mongers should be considered in future work. Two possible directions are summation of the probabilities yielded by the hate post classifier, and averaging the number of hate posts per user are two optional alternatives. 
However, we wish to stress that opting for a conservative labeling of seed users achieves a cleaner diffusion process -- a process that is usually prone to noise.

\paragraph{Parler users} In this work, we make use of a Parler dataset introduces by \citet{aliapoulios2021early}. One limitation of this dataset is that it includes only part of the Parler full corpus. However, data were not sampled at random --  the authors retrieved data based on users' identity (i.e., all data for 4.08M users out of 13.25M), providing a decent coverage of a significant part of the network. 

\paragraph{Time span} Given that we provide a comparison between  trends in Parler and Gab, it is important to note the datasets span different and non-overlapping time-frames (see Table \ref{table:Data_statistics}). Therefor, the comparison we provide should be read cautiously. We do note, however, that each of the datasets was crawled from the early days of the social platform and spans over a similar range of time (17 months). Moreover, the time disparity between the dataset could be considered as an advantage -- allowing to examine the generalization performance of hate speech models, as we report in Section \ref{subsec:results_post_level}.

\ignore{
    \paragraph{The $\widetilde{HM}$ Population} Users that belong to the $\widetilde{HM}$ population are not significantly hate mongers. However, they are not clear normal users. We notice that, in most of our analysis in Parler and Gab, the $\widetilde{HM}$ users are most similar to the $N$ population. We suspect that there are two central reasons why these $\widetilde{HM}$ users are not fully defined as $HM$: (i) Parler social network has a relatively low density. The average clustering coefficient \cite{albert2002statistical} in Parler is 0.153 while in Gab the average is significantly higher (0.25)  -- limiting the hate to ``easily propagate'' during the diffusion process (ii) Parler social network contains only a partial set of the nodes (users). It creates a higher amount of connected components which restricts the diffusion process between the different components. However, in Gab we see the opposite effect. $\widetilde{HM}$ users are most similar to the $N$ users.
}

\paragraph{Ethical Considerations} Analysing and modeling hate speech in a new social platform such as Parler is of great importance. However, classifying \emph{users} as hate mongers, based on the output of an algorithm, may result in marking users falsely (which may result in suspension or other measures taken against them). While we always opted for a conservative approach, as well as focusing on aggregate measures characterizing the trends of a \emph{platform}, we note that user labeling should be used in a careful manner, ideally involving a `man-in-the-loop'.

Considering the annotation task -- the annotation process did not include any information about the identity of the users. In addition, we warned our human annotators about the possible inappropriate content of the posts.

\section{Conclusion}
\label{sec:summary_and_future_work}
To the best of our knowledge, we present the first large-scale computational analysis of hate speech on Parler, and provide a comparison to trends observed in the Gab platform.

We annotate and share a the first Parler dataset, containing 10K posts labeled by the level of hate they convey. We used this dataset to fine-tune a transformer model to be used to mark a seed set of users in a diffusion model, resulting in user-level classification. 
We find significant differences between hate mongers ($HM$) and other user groups: $HM$ represent only 16.1\% and 10\% of the active users in Parler and Gab respectively. However, they create \emph{41.23\%} of the content in Parler and \emph{71.38\%} of the content in Gab. We find that $HM$ are show higher engagement and they have significantly more followers and followees.
Other differences are manifested through the sentiment level expressed and the emotions conveyed.



\bibliography{hateful_users_bib.bib}
\end{document}